\documentclass[useAMS,usenatbib,usegraphicx]{mn2e}
\usepackage{times}

\begin{document}

\newcommand{\be}{\begin{equation}}
\newcommand{\ee}{\end{equation}}
\newcommand{\ts}{\textstyle}
\newcommand{\ds}{d_{\rm s}}
\newcommand{\vt}{v_{\rm t}}
\newcommand{\vl}{v_{\rm l}}
\newcommand{\thete}{\theta_{\rm E}}
\newcommand{\tv}{t_{\rm v}}
\newcommand{\te}{t_{\rm E}}
\newcommand{\tfw}{t_{\rm 1/2}}
\newcommand{\ut}{u_{\rm t}}
\newcommand{\sm}{{\rm M}_{\sun}}

\title[The Angstrom Project]{The Angstrom Project: a microlensing survey
of the structure and composition of the bulge of the Andromeda galaxy}
\author[Kerins et al.]{E.~Kerins,$^1$\thanks{Email: ejk@astro.livjm.ac.uk} M.J.~Darnley,$^1$ J.~Duke,$^1$ A.~Gould,$^2$ C.~Han,$^3$ Y.-B. Jeon,$^4$ A.~Newsam$^1$ \newauthor
and B.-G.~Park$^4$ (The Angstrom Collaboration)\\
$^1$Astrophysics Research Institute, Liverpool John Moores
 University, Twelve Quays House, Birkenhead, Merseyside CH41 1LD\\
$^2$Department of Astronomy, Ohio State University, 140 West 18th Avenue,
Columbus, OH 43210, USA\\
$^3$Department of Physics, Institute for Basic Science Researches, 
Chungbuk National University, Chongju 361-763, Korea\\
$^4$Korea Astronomy and Space Science Institute, 61-1,Whaam-Dong, Youseong-Gu, Daejeon 305-348, Korea
}
\maketitle

\begin{abstract}
The Andromeda Galaxy Stellar Robotic Microlensing Project (The Angstrom Project)
aims to use stellar microlensing events to trace the structure and composition
of the inner regions of the Andromeda Galaxy (M31). We present microlensing
rate and timescale predictions and spatial distributions for
stellar and
sub-stellar lens populations in combined disk and barred bulge models of M31. We
show that at least half of the stellar microlenses in and around the bulge are
expected to have
characteristic durations between 1 and 10 days, rising to as much as 80\% for
brown-dwarf dominated mass functions. These short-duration
events are mostly missed by current microlensing surveys that are looking for
Macho candidates in the M31 dark matter halo. Our models predict that an
intensive monitoring survey programme such as Angstrom, which will be able to
detect events of durations upwards of a day, could detect around 30 events per
season
within $\sim 5$~arcminutes of the M31 centre, due to ordinary low-mass stars and
remnants. This yield increases to more than 60 events for brown-dwarf dominated
mass
functions. The overall number of events and their average duration are
sensitive diagnostics of the bulge mass, in particular the
contribution of low-mass stars and brown dwarfs. The combination of an inclined
disk,
an offset bar-like bulge, and differences in the bulge and disk luminosity
functions results in a
four-way asymmetry in the number of events expected
in each quadrant
defined by the M31 disk axes. The asymmetry is sensitive to
the bar prolongation, orientation and mass.
\end{abstract}

\begin{keywords}
gravitational lensing -- stars: low mass, brown dwarfs -- galaxies: bulges
-- galaxies: individual: M31 -- galaxies: structure 
\end{keywords}

\section{Introduction} \label{intro}

Over the last decade the microlensing effect has been used to constrain the
abundance of compact dark matter in the Milky Way halo
\citep[e.g.][]{alc00a,afo03a}. However, some of the most spectacular results
have come not from the dark matter surveys but from stellar microlensing surveys
directed towards the Galactic bulge, where thousands of events involving
ordinary stars have been detected \citep[e.g.][]{alc00b,uda03,afo03b,sumi03}.
The optical depth measured by these surveys has shown that the Milky Way bulge
is bar-like and oriented almost towards us, a conclusion supported by
near-infrared surveys of the bulge \citep{dwek95}. There remains some spread in
the reported optical depth of the Milky Way bulge by the different survey teams,
and this uncertainty partly reflects the difficulty involved in separating
foreground disk and bulge lensing populations, as well as the fact that the
number of lines of sight through the bulge is severely restricted due to dust
obscuration. Follow-up survey teams, which monitor events alerted in real-time
by
the main survey teams, have also obtained high signal-to-noise ratio lightcurves
for many events, including the spectacular binary event EROS~BLG-2000-5, which
allowed an accurate mass determination to be made of the lens \citep{an02}, as
well as a detailed limb-darkening profile of the source \citep{fie03}. These
surveys have also produced the first microlensing detection of a planet
\citep{bond04}. Microlensing in our own galaxy is now an established tool for
planetary, stellar and galactic astrophysics.

\begin{figure}
\includegraphics[scale=0.35,angle=0]{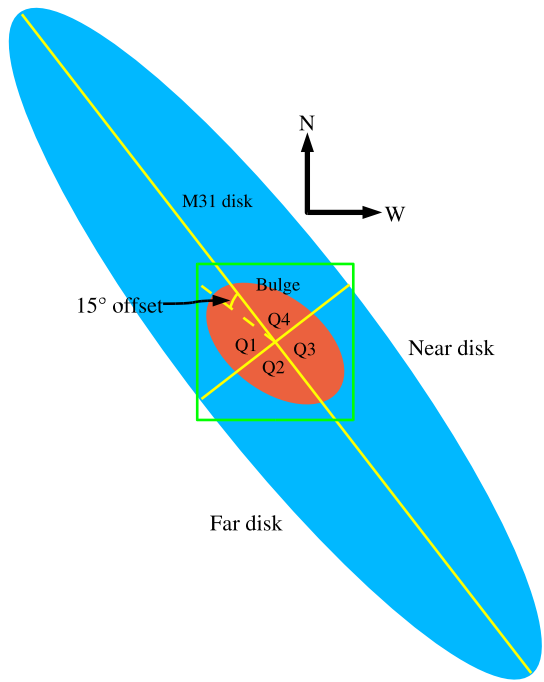}
\caption{The geometry of the M31 disk and bulge. The disk of M31 is inclined at an angle of $77\degr$, with the north-western side nearer to us. The bulge major axis is taken
to be offset on the sky from the
disk major axis by $15\degr$, as indicated by the twisting of the inner surface
brightness isophotes \citep{star94}.
The M31 disk major and minor axes define 4 quadrants, labelled Q1-4, which are
referred to later in this paper.
Events involving bulge lenses are expected to occur mainly in quadrants Q1 and
Q2 where the disk stars lie behind the bulk of the bulge stars, whilst events
involving disk lenses preferentially occur in quadrants Q3 and Q4 where they are
in front of the bulk of the bulge stars.}
\label{geom}
\end{figure}

With the advent of techniques like difference imaging \citep{ala98} it has
become possible to look for microlensing events in other galaxies where the
sources are unresolved, the so-called pixel-lensing regime. Several teams
\citep{pau03,cal03,rif03,dej04,ugle04,jos05} are already exploiting this and similar
techniques to look for Macho dark matter candidates in the halo of the Andromeda
Galaxy (M31). In this paper we consider how an intensive stellar pixel-lensing
survey of the bulge of M31 can be used to probe the stellar mass function and
the inner M31 galactic structure. In some respects we have a better view of the
M31
bulge than of our own. The M31 disk is inclined at $77\degr$, so we have a
relatively unobscured view of the bulge. M31 is believed to be of earlier type
than our own galaxy, and therefore thought to possess a more prominent bulge
component. Intriguingly, surface brightness measurements of the central regions
of
M31 reveal a twisting of the inner optical isophotes \citep{wal87}, consistent
with a
bar-like bulge that is misaligned from the disk major axis by about $15\degr$
on the sky \citep{star94}. The relationship between the M31 disk and bulge is
shown schematically in Figure~\ref{geom}.

The Andromeda Galaxy Stellar Robotic Microlensing (Angstrom) Project aims
to detect stellar microlensing events in the bulge region of M31. By
obtaining upwards of three epochs per 24-hour period the survey will be
sensitive to short-duration stellar microlensing events due to low mass stars
and brown dwarfs, which are missed by current M31 microlensing surveys looking
for
dark matter. Angstrom is using three telescopes in the Northern hemisphere at
widely separated longitudes. A pilot season has already begun with the robotic
2m Liverpool Telescope on La~Palma, the 1.8m Doyak telescope at the Bohyunsan Observatory in Korea and the
2.4m Hiltner at the MDM Observatory in Arizona. This paper considers how many
events a survey like Angstrom should see, and how their number, timescales and
spatial distribution are sensitive to the low-mass end of the stellar mass
function,  and to the bulge mass and geometry. Previous studies
\citep[e.g.][]{gou94,han96} have considered the optical depth and event rate
from simple disk and spheroidal bulge populations. In this study we make
a detailed examination of the pixel-lensing signal expected from a barred
bulge, based on the survey sensitivity of the Angstrom Project. 

The format of the paper is as follows. In Section~\ref{theory} we discuss the
basics of pixel-lensing theory and derive an expresion for the pixel-lensing
rate for simple analytic distribution functions. In Section~\ref{simpmodels} we
construct a set
of five stellar microlensing models comprising a disk and barred bulge. The
models
reproduce the observed M31 surface brightness and are carefully constructed
using consistent assumptions regarding the stellar mass function, luminosity
function and resulting mass-to-light ratio. The microlensing predictions from
the models are presented in Section~\ref{predictions} and we end with a
discussion of the results in Section~\ref{discuss}.

\section{Pixel-lensing theory} \label{theory}

Pixel lensing differs from classical microlensing in that the sources are
unresolved at baseline. Pixel-lensing events are observed as transient local
flux enhancements against the background surface brightness. An event
is detectable when the lens approaches sufficiently close to the line of sight
to an unresolved source. When the lens passes within a fraction $u$ of its
Einstein radius the background source becomes magnified by a factor $A =
(u^2+2)/[u^2(u^2+4)]^{1/2}$, which approximates to $u^{-1}$ for the
high-magnification events that comprise the large majority of 
pixel-lensing lightcurves. A detection is possible when the excess
photon count $(A-1)N_*$ due to the microlensed source star exceeds the local
noise, which for bright regions of galaxies is dominated by the galaxy count
$N_{\rm gal}$. For purely Poisson
noise we require
   \be
      (A-1)N_* \simeq u^{-1}N_* > \alpha N_{\rm gal}^{1/2}, \label{detect}
   \ee
where $\alpha = 3$ is a typical theorists choice. Let us define $\ut$
to be the threshold impact parameter whereby events with a minimum impact
parameter $u_0 < \ut$ satisfy equation~(\ref{detect}) and so are
detectable. These events are detectable for a timescale
   \be
      \tv = 2  (\ut^2 - u_0^2)^{1/2} \frac{\theta_{\rm E}}{\mu} = 2  (\ut^2 -
      u_0^2)^{1/2} \te, \label{tf}
   \ee
with
   \be
       \thete = \left[    \frac{4 G m}{c^2\ds} (l^{-1}-1) \right]^{1/2}
    \ee
the angular size of the Einstein radius, $m$ the lens mass, $\ds$
 the distance to the source star and $l$ the distance to the lens in units
of $\ds$. Here $G$ and
$c$ denote the standard physical constants. In equation~(\ref{tf}) $\mu$ is
the relative proper motion of the lens across the line of sight and $\te =
\theta_{\rm E}/\mu$ is the Einstein radius crossing time. We call $\tv$ the
{\em visibility timescale} of the pixel-lensing event. To characterise an event
we
must sample with a frequency that is much higher than $\tv^{-1}$. In this case
each of our observations of the event around peak has a significance of {\em at
least} $\alpha$ and the significance of the event detection as a whole is much
greater.

The differential rate of pixel microlensing for lenses of mass $m$ against a
given source star can be
expressed as 
   \be
	d^5\Gamma_m = \frac{\ut}{m} F \ds^5 l^4 \mu^2 \thete
	\, d\vl \, d\mu \, du \, d\phi \, dl, \label{dr}
   \ee
where $F$ is the lens distribution function,
$\vl$ is the lens velocity along the
line of sight, $u < \ut$ is the lens impact parameter, $\psi$ is the lens
mass function and $\phi$ describes the
random orientation of the lens and source on the sky. Using equation~(\ref{tf})
to change
variables from $\mu$
to $\tv$ we can re-write equation~(\ref{dr}) as
	\be
		d^5\Gamma_m  =  \frac{8}{m} \ut (\ut^2 - u^2)^{3/2} F \ds
		\left(	\frac{\thete \ds l}{\tv} \right)^4  
                \, d\vl \, d\tv
		\, du \, d\phi	\, dl, \label{drate}
	\ee
from which the event timescale distribution $d\Gamma_m/d\tv$ can be obtained.

For the models discussed in this paper we shall assume simple distribution
functions comprising an analytic density function $\rho$ and an isotropic 
Maxwellian velocity distribution. That is
   \be
      F = \frac{\rho}{(2\pi)^{3/2} \sigma^3} \exp \left[-\frac{(\vt^2 +
\vl^2)}{2 \sigma^2} \right ], \label{df}
   \ee
where $\rho$ has a functional form yet to be
specified, $\vt$ is the velocity across the line of
sight and $\sigma$ is the one-dimensional velocity dispersion.
Equation~(\ref{drate}) can then be integrated to give the timescale
distribution
\begin{eqnarray}
    \frac{d\Gamma_m}{d\tv} & = & \frac{32 \, \sigma}{m} \int_0^1 \int_0^{\ut} 
    \frac{\ut \rho(l)
    \gamma^4}{(\ut^2-u^2)^{1/2}} \exp(-4\gamma^2) \nonumber \\
    & &  \quad \quad \quad\quad \quad \quad \quad \quad 
    \times I_0(4 \gamma \delta)
    \exp(-\delta^2) \,dl\, du, \label{timedis}
\end{eqnarray}
where $I_0(x)$ is the zeroth-order modified Bessel function, $\gamma = \thete l
\ds (\ut^2-u_0^2)^{1/2}/\sqrt{2} \sigma \tv$ and $\delta = \mu_{\rm sys} l
\ds/\sqrt{2} \sigma$, with $\mu_{\rm sys}$ the proper motion of the line of
sight. When considering a range of lens masses we must integrate the timescale
distribution over the lens mass function $\psi(m)$:
   \be
      \frac{d\Gamma}{d\tv} = \frac{\int_{m_{\rm l}}^{m_{\rm u}}
(d\Gamma_m/d\tv ) \, \psi(m) \, dm}{\int_{m_{\rm l}}^{m_{\rm u}} \psi(m) \, dm},
\label{massint}
   \ee
where $m_{\rm l}$ and $m_{\rm u}$ are the lower and upper lens mass cut-offs,
respectively. The
integral of $d\Gamma/d\tv$ over $\tv$ then provides the pixel-lensing event
rate
per source, $\Gamma$.

From equation~(\ref{detect}) we can see that $\ut$ depends
upon both the source
luminosity and the background surface brightness. Moreover, $d\Gamma /d \tv$
is also sensitive to the source distance $\ds$. Therefore a full evaluation of
$d\Gamma/d\tv$ at a single
timescale $\tv$ and along a single line of sight involves solving a
five-dimensional integral over $l$,
$\ds$, $u$, $m$ and source luminosity. An evaluation of $\Gamma$  over
the whole bulge region requires two further integrals over $\tv$ and galaxy
surface brightness.
We evaluate the
integrals over $l$, $\ds$, $\tv$ and surface brightness via direct numerical
integration, whilst the integrals over $u$, $m$ and source luminosity are
performed using Monte-Carlo techniques.

A further complication is that equation~(\ref{detect}) and subsequent formulae
strictly hold only if the source star can be assumed to be point-like, that is
to say its angular size $\theta_* \ll u \thete$. This approximation may break
down when bright giant stars are microlensed by low-mass brown dwarfs in
the disk or bulge. In this finite-source regime there is differential
magnification across the face of the source star and so the overall
magnification is evaluated by integrating the microlensing magnification over
the source. In our calculations we explicitly take account of finite source
effects when necessary.

\section{Simple models for the M31 inner galaxy} \label{simpmodels}

Microlensing predictions necessarily depend upon the choice of model one adopts
for the lens and source populations. There
is a good deal of uncertainty in  some of the key parameters, which in itself
presents one of the prime drivers for the Angstrom Project. Given this
uncertainty, it is sufficient for us to adopt relatively simple 
models for the M31 stellar population that contain the key parameters we are
interested in exploring. To this end we consider a simple two-component model
for the stellar populations in M31 comprising an exponential disk and a barred
bulge. 

At first sight stellar pixel lensing predictions appear to be at the mercy of
several theoretical distributions, many of which are ill-constrained
or unknown for M31. These include (for both the disk and bulge components) the
density and velocity distributions, the stellar and remnant mass functions,
and the source luminosity functions. However these distributions are tightly
linked with one another and we can exploit this in order to reduce the parameter
space. Firstly, we demand that our bulge and disk models are able to
satisfactorily reproduce the observed M31 surface brightness profile along both
the major and minor axes. Secondly, we demand that our assumed luminosity and
mass functions are consistent with one another, that is to say our assumed
luminosity functions fix the shape of the upper end of our mass functions where
the bulk of the light is produced. Thirdly, the low mass end of our mass
functions fix the density normalisations for the disk and bulge and thereby
their mass-to-light ratios, since the light is fixed by the observed surface
brightness profile.

We proceed by investigating ``heavy'' and ``light'' disk and bulge models. The
heavy models assume stellar mass functions that extend into the brown dwarf
regime and that contain a significant density in such objects. The light
models comprise just ordinary hydrogen burning stars with a mass distribution
similar to Milky Way stellar populations. All disk and bulge models
also comprise a stellar remnant contribution and the disk models are 
assumed to comprise a significant amount of gas which neither provides sources
nor lenses for microlensing. The mass-to-light ratio implied by the adopted mass
functions, along with the gas and remnant contributions, sets the density
normalisation for each component by the requirement that the combined light
from the disk and bulge satisfactorily reproduces the M31 surface brightness
profile along the major and minor axes.

\subsection{Density and velocity distributions} \label{functions}
 
Our first component is the disk, for which we adopt a conventional
double-exponential profile:
   \be
      \rho_{\rm d} = \rho_{{\rm d},0} \exp (-R/h) \exp (-|z|/H) 
      \label{diskden} 
   \ee
where $\rho_{{\rm d},0}$ is the central disk density, $h = 5.8$~kpc
is the disk scale-length, $H = 0.4$~kpc is the scale-height, and $R$ and
$z$ are cylindrical coordinates. For our heavy disk we find $\rho_{{\rm d},0} =
0.3~\sm$~pc$^{-3}$ gives an $R$-band mass-to-light ratio $M/L_R = 3$
and reproduces well the M31 surface brightness away from the bulge, whilst for
the light disk model we set the normalisation a factor 3.4 times lower.

Disk lenses and sources are assumed to orbit the disk with a circular speed of
235~km~s$^{-1}$ beyond 1.25~kpc (5~arcminutes) of the M31 centre. Inside
1.25~kpc the disk is taken to be a solid body rotator. Stars in the
heavy disk are
assumed to have random motions described by the Maxwellian velocity
distribution in equation~(\ref{df}), with a mean one-dimensional
dispersion in the transverse direction of $\sigma =
60$~km~s$^{-1}$ at $R =
2 h$, a position roughly equivalent to the location of the Sun in the Milky Way;
for the light  disk we assume a dispersion that is a factor $\sqrt{3}$ lower
than this. For
other cylindrical distances $R$ the velocity dispersion is scaled by
$\exp(-R/2h)$, that is by the square root of the mid-plane density.

The second component is the bulge. The twisting of the optical isophotes in the
inner regions of M31 clearly indicates that the M31 bulge is barred and with a
major axis which is offset by some $15\degr$ on the sky from the M31 disk
major axis \citep{star94}, as illustrated schematically in Figure~\ref{geom}.
The surface
brightness profile of galactic bulges and elliptical galaxies is
typically observed to follow the de~Vaucouleur exponential $R^{1/4}$ law
\citep{wyse97}.
However, in the inner regions a power-law profile with surface
brightness $I
\propto R^{-\alpha}$ is often observed \citep{bin98}, implying a volume density
fall-off $\rho
\propto R^{-(1+\alpha)}$. We shall consider two possible profiles, one an
exponential and
one a power law. Both distributions are modelled as bars that are offset from
the M31 disk major axis. 

\begin{figure}
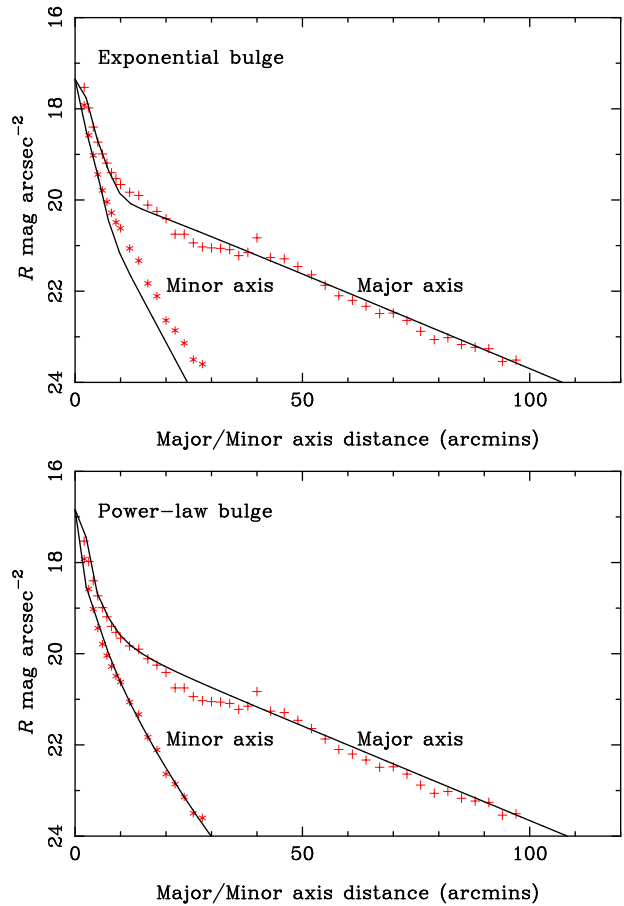

\includegraphics[scale=0.35,angle=270]{sb-exp.ps}
\includegraphics[scale=0.35,angle=270]{sb-power.ps}
\caption{Major and minor axis surface brightness profiles predicted for models
1--4 involving an exponential bulge (top) and for model 5 involving a power-law
bulge (bottom). The model fits are shown by the solid lines whilst the crosses
are $R$-band data for the South-Western major axis and South-Eastern minor axis
from \citet{wal87}. The exponential bulge model fit is evidently
poor beyond 10~arcminutes along the minor axis, but is a good fit within the
inner
bulge region of interest in this paper.}
\label{sb}
\end{figure}

For the exponential bulge we adopt the following
distribution:
  \be
      \rho_{\rm b} = \rho_{{\rm b},0} \exp \{-[(x_{\rm b}/a)^2+(y_{\rm b}/q
      a)^2 +(z_{\rm b}/q a)^2]^s\},
      \label{blgden1}
   \ee
where  $a = 1$~kpc is
the bulge scale length, $s = 0.75$ is a power-law index, $q = 0.6$ defines the
bar prolongation and $x_{\rm b}$,
$y_{\rm b}$ and $z_{\rm b}$ are
Cartesian coordinates aligned along the bar principal axes, with $z_{\rm b}$
normal to the disk plane.  The choice of $a$ and $s$ provides a good fit to
the surface brightness profile within 10~arcminutes, the region in which we
are interested, though the fit becomes poorer at larger radii (see top panel of
Figure~\ref{sb}). For the heavy bulge we fix $\rho_{{\rm b},0} =
12~\sm$~pc$^{-3}$
giving $M/L_R = 7$ and a bulge mass of $3\times 10^{10}~\sm$, which is 
towards the upper end of favoured mass estimates \citep[c.f.][]{ken89,wid03}.
For the light
bulge
the density is 3.4 times lower, which is at the lower
end of the range of
mass estimates and is comparable to the inferred mass of the
Milky Way bulge \citep{dwek95}. 

For the power-law distribution we
take
   \be
      \rho_{\rm b} = \rho_{{\rm b},0} \{1+[(x_{\rm b}/a)^2+(y_{\rm b}/q a)^2
+(z_{\rm b}/q a)^2]\}^{-s/2}.
      \label{blgden2}
   \ee
Here we adopt $a = 0.75$~kpc, $q = 0.6$ and a power-law index $s =
3.5$. The parameters $s$ and $a$ are chosen to allow a
good
match to the shape of the observed major- and minor-axis surface brightness
profiles (bottom panel of Figure~\ref{sb}).. The density normalisation for the
power-law bulge is fixed
at $\rho_{{\rm b},0} =
6~\sm$~pc$^{-3}$, giving a total mass of $1.7 \times 10^{10}~\sm$,
intermediate to the heavy and light exponential bulge, though  $M/L_R
= 2.1$ which is the same as for the light exponential bulge.

\begin{table*}
\begin{minipage}{160mm}
\caption{The M31 disk-bulge models. Parameters for the density and velocity
refer to the density distributions in section~\ref{functions} and the
distribution function in equation~(\ref{df}). Parameters for the lens masses
refer to the mass functions discussed in section~\ref{masses}. For the disk the
quoted velocity dispersion is normalised to its value at cylindrical distance $R
= 2h$. Other parameters which are fixed to the same values for all models are
discussed in sections~\ref{functions} and \ref{masses}.}
\label{models}
\begin{tabular}{@{}lllll}
\hline
Model & Description & Bulge parameters & Disk parameters & Remarks \\
\hline
1 & Light exponential & $\rho_{{\rm b},0} = 3.5~\sm$~pc$^{-3}$, $\sigma =
90$~km~s$^{-1}$ & $\rho_{{\rm d},0} = 0.09~\sm$~pc$^{-3}$, $\sigma =
35$~km~s$^{-1}$ & \\
  & bulge, light disk & $m_{\rm l} = 0.08~\sm$, $x_1 = -1.4$ & $m_{\rm l} =
0.08~\sm$, $x_1 = -1.4$ & \\
  & & $a = 1$~kpc, $M/L_R = 2.1$ & $M/L_R = 0.9$ &  \\
2 & Heavy exponential & $\rho_{{\rm b},0} = 12~\sm$~pc$^{-3}$, $\sigma =
150$~km~s$^{-1}$ & $\rho_{{\rm d},0} = 0.3~\sm$~pc$^{-3}$, $\sigma =
60$~km~s$^{-1}$ & \\
  & bulge, heavy disk & $m_{\rm l} = 0.03~\sm$, $x_1 = -2.35$ & $m_{\rm l} =
0.01~\sm$, $x_1 = -2.35$ & \\
  & & $a = 1$~kpc, $M/L_R = 7$ & $M/L_R = 3$ &  \\
3 & Light exponential & $\rho_{{\rm b},0} = 3.5~\sm$~pc$^{-3}$, $\sigma =
90$~km~s$^{-1}$ & $\rho_{{\rm d},0} = 0.3~\sm$~pc$^{-3}$, $\sigma =
60$~km~s$^{-1}$ & \\
  & bulge, heavy disk & $m_{\rm l} = 0.08~\sm$, $x_1 = -1.4$ & $m_{\rm l} =
0.01~\sm$, $x_1 = -2.35$ & \\
  & & $a = 1$~kpc, $M/L_R = 2.1$ & $M/L_R = 3$ &  \\
4 & Light exponential & $\rho_{{\rm b},0} = 4~\sm$~pc$^{-3}$, $\sigma =
90$~km~s$^{-1}$ & $\rho_{{\rm d},0} = 0.3~\sm$~pc$^{-3}$, $\sigma =
60$~km~s$^{-1}$ & Uses disk LF and \\
  & bulge, heavy disk & $m_{\rm l} = 0.01~\sm$, $x_1 = -2.35$ & $m_{\rm l} =
0.01~\sm$, $x_1 = -2.35$ & MF for both bulge\\
  & & $a = 1$~kpc, $M/L_R = 2.4$ & $M/L_R = 3$ & and disk \\
5 & Power-law bulge, & $\rho_{{\rm b},0} = 6~\sm$~pc$^{-3}$, $\sigma =
120$~km~s$^{-1}$ & $\rho_{{\rm d},0} = 0.09~\sm$~pc$^{-3}$, $\sigma =
35$~km~s$^{-1}$ & \\
  & light disk & $m_{\rm l} = 0.08~\sm$, $x_1 = -1.4$ & $m_{\rm l} = 0.08~\sm$,
$x_1 = -1.4$ & \\
  & & $a = 0.75$~kpc, $M/L_R = 2.1$ & $M/L_R = 0.9$ &  \\
\hline
\end{tabular}
\end{minipage}
\end{table*}

The $R$-band surface brightness distribution predicted by the combined disk and
exponential bulge model is shown in the upper panel of Figure~\ref{sb}, whilst
the lower panel shows the combined disk and power-law bulge profile.
Also shown are $R$-band surface brightness measurements along the major and
minor axis of the M31 disk made by \citet{wal87}.
The distributions in
equations~(\ref{blgden1}) and (\ref{blgden2}) differ dramatically at large
distances from the bulge centre. The exponential profile gives a considerably
poorer fit
than does the power-law profile at distances beyond 10~arcminutes along the M31
minor axis. However, the two distributions are necessarily similar at
smaller distances within the region
of interest to us, with the power-law bulge having a slightly brighter peak.

The bulge stars are taken to have Maxwellian random motions according to
equation~(\ref{df}) with $\sigma = 90$~km~s$^{-1}$ for the light exponential
bulge, a factor $\sqrt{2}$ higher for the power-law bulge and a factor
$\sqrt{3}$ higher for the heavy exponential bulge. The dispersions are
consistent with the range of $\sigma = 100 - 150$~km~s$^{-1}$ observed by
\citet{mcel83}. The bulge velocity dispersion could well differ significantly
along each of the principal axes of the bar, so our chosen $\sigma$ represents
the square root of the mean two-dimensional transverse dispersion. We also adopt
a bar pattern
speed of 57~km~s$^{-1}$~kpc$^{-1}$ out to 3.2~kpc \citep{star94},
capping the
circular speed at 180~km~s$^{-1}$ outside of this.

\subsection{Source luminosities} \label{lf}

From equation~(\ref{detect}) the intrinsic luminosity of the sources is a key
factor in detecting pixel-lensing events. Fainter sources are more numerous but
require higher magnification in order to be detected. The expected rate of pixel
lensing
therefore depends on the luminosity function of the sources, but tends to be
dominated by the brighter stars. 

The luminosity function of the M31 bulge region has been explored in several
studies \citep[e.g.][]{dav91,rich95,jab99,ste03}. The high stellar densities
cause severe blending problems, which hamper accurate calibrations of the
bright end of the stellar luminosity function. In so far as
this can be corrected, recent studies find no significant evidence for
differences between the bright end of the stellar luminosity function in the M31
and Milky Way bulge regions \citep{ste03}.

\begin{figure}
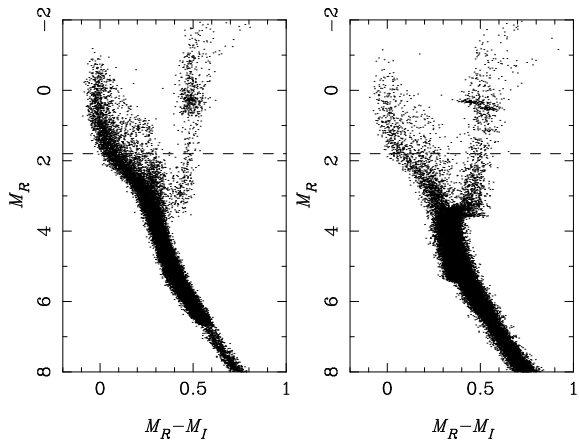

\includegraphics[scale=0.25,angle=0]{disk-stars.ps}
\includegraphics[scale=0.25,angle=0]{bulge-stars.ps}
\caption{The disk (left) and bulge (right) source star synthetic $R$
versus $R-I$ colour magnitude distributions used for our pixel-lensing
calculations. For the Angstrom Project stars above the dashed lines
at $M_R = 1.8$ are most likely to act as sources to pixel-lensing events as they
require a magnification $A < 30$ to satisfy equation~(\ref{detect}) for a
background surface brightness $\mu_R = 20$~mag~arcsec$^{-2}$.}
\label{sources}
\end{figure}

We use synthetic colour-magnitude data to model the M31 source stars.
These have been generated to match
the chemical composition and the mass and luminosity functions observed for
stars in the Milky Way disk and bulge and we assume they provide an adequate
description of stars in the M31 disk and bulge. The synthetic datasets are based
upon the theoretical stellar isochrones of the Padova group \citep{gir01}. The
disk and
bulge datasets each comprise around a million simulated stars with tabulated
masses, ages, bolometric luminosities, effective temperatures and
absolute magnitudes in
several passbands. The $R$ versus $R-I$ colour--magnitude diagrams for a subset
of the disk and bulge stars are shown in Figure~\ref{sources}.

The main advantage in using synthetic datasets for microlensing
calculations is that the radius of each star can be readily computed and so we
can allow for finite source effects in a consistent way. For all but one
of our microlensing models we shall assume the M31 bulge sources are described
by the synthetic Milky
Way bulge stars and the M31 disk sources by the synthetic Milky Way disk stars.
For the other model we assume both bulge and disk sources are described by the
synthetic Milky Way disk stars. This allows us to assess the impact of our
choice of luminosity function upon the pixel lensing rate predictions.

We allow for for a combined foreground and internal M31 $R$-band extinction of $A_R = 0.5$~mag 
when computing the mass-to-light ratio of the source stars. This is equivalent to three times the foreground value \citep{sch98} and is consistent with extinction measurements of Sb galaxies using background galaxy counts \citep{hol05}.

\begin{figure*}
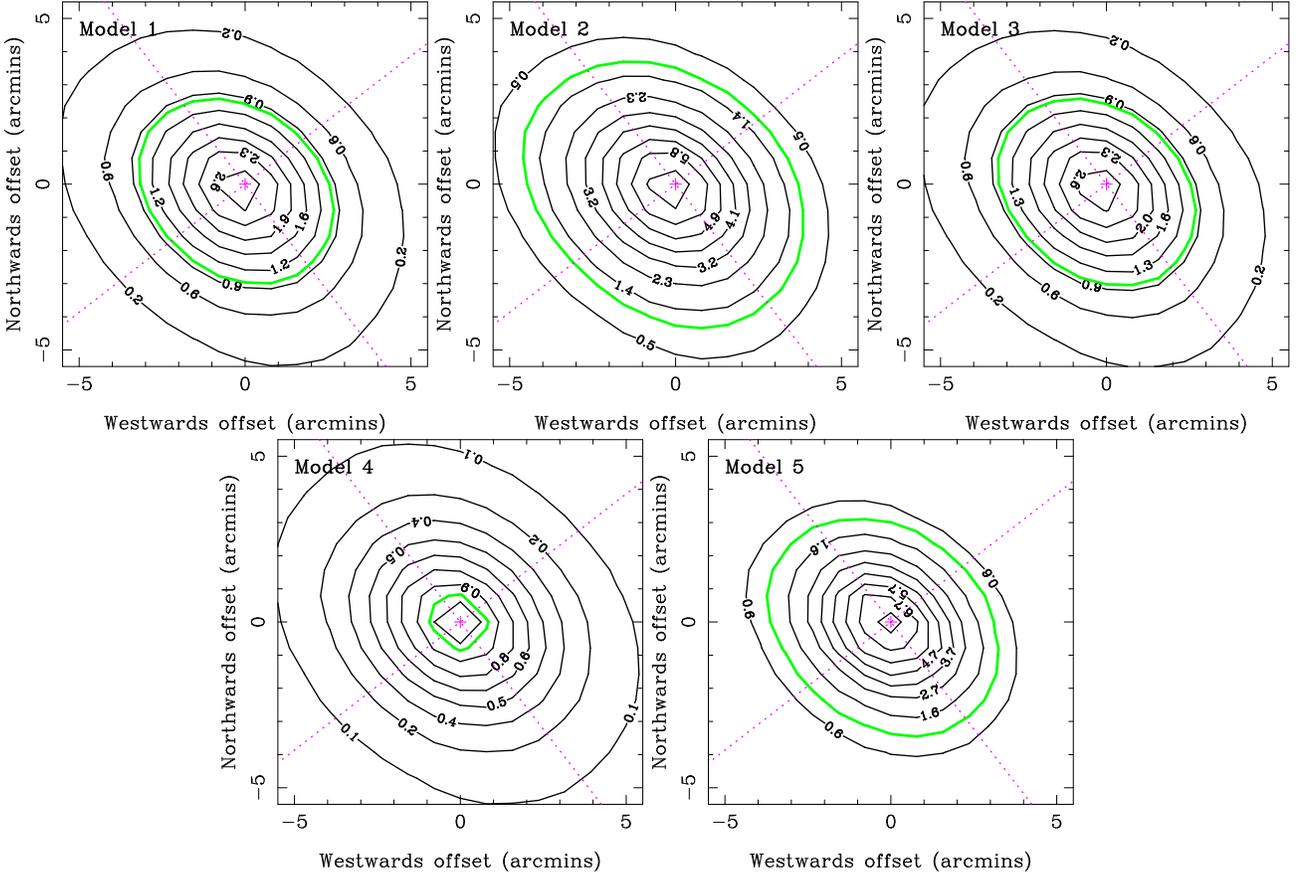

\includegraphics[scale=0.35,angle=270]{spatial-model1.ps}
\includegraphics[scale=0.35,angle=270]{spatial-model2.ps}
\includegraphics[scale=0.35,angle=270]{spatial-model3.ps}
\includegraphics[scale=0.35,angle=270]{spatial-model4.ps}
\includegraphics[scale=0.35,angle=270]{spatial-model5.ps}
\caption{The pixel lensing spatial distribution for models 1--5 for events with
visibility timescales $\tv = 1$ -- 100 days. The origin of each panel
corresponds to the centre of M31 (shown by the cross) and the dotted lines
running diagonally demarcate the major (upper left to lower right) and minor
(lower left to upper right) axes of the M31 disk. The event rate contours are
shown in black and are labelled in events per year per arcminute$^2$. The green
contour indicates a level of 1 event per year per arcminute$^2$.}
\label{spatial}
\end{figure*}

\subsection{Stellar and remnant mass functions} \label{masses}

For our light models we assume that the bulge and disk mass functions in M31
are similar to those of the Milky Way disk and bulge but do not extend into the
brown dwarf regime. We assume a mass function
of the form $\psi(m) = \psi_* + \psi_{\rm rem}$, where $\psi_*$ and $\psi_{\rm
rem}$ are the mass functions of the stars and remnants, respectively. The
remnant mass function is taken to be a superposition of three Dirac delta
functions peaked at 0.6, 1.35 and $5~\sm$, corresponding to white dwarfs,
neutron stars and black holes, respectively. The ratio of their mass
density contributions is set at 0.7:0.2:0.1, respectively. This prescription is
a slight simplification of the remnant mass function advocated by \citet{gou00}.

The stellar mass function $\psi_*$ is modelled by a broken power law:
   \be
       \psi_* = \left\{ \begin{array}{ll}
       K (m/m_{\rm t})^{x_1} & (m_{\rm l} < m < m_{\rm t} ) \nonumber \\
       K (m/m_{\rm t})^{x_2} & (m_{\rm t} \leq m < m_{\rm u} )
       \end{array} \right.
   \ee
between lower and upper mass limits $m_{\rm l}$ and $m_{\rm u}$. We set the
transition mass at $m_{\rm t} = 0.5~\sm$.  The power law
index $x_2$ and $m_{\rm u}$ are fixed to be consistent with our synthetic
stellar population models. Accordingly we take $m_{\rm u} = 1~\sm$ for bulge
stars and $10~\sm$ for disk stars, whilst $x_2 = -2.35$ for both populations.
This is the case for both the heavy and light models. For the light bulge and
disk we take $m_{\rm l} = 0.08~\sm$ and $x_1 = -1.4$, which is consistent with 
observations of the Milky Way bulge mass function \citep{zoc00}. For
the
heavy models we instead let $x_1 = x_2 = -2.35$ and extend the mass function
down to $m_{\rm l} = 0.03~\sm$ for the bulge and $m_{\rm l} = 0.01~\sm$ for the
disk. Different values of $m_{\rm l}$ are needed in order to reproduce the
correct $M/L_R$ values for each component, though we could just as easily have
used a single value of $m_{\rm l}$ and different values for $x_1$.

The normalisation constant $K$ is fixed by the overall stellar mass
density, which contributes a fraction $(1-f_{\rm rem}-f_{\rm gas})$ of the
overall density, where $f_{\rm rem}$ and $f_{\rm gas}$ are the remnant and gas
density contributions, respectively. We assume $f_{\rm gas} = 0$ in the bulge
and 0.3 in the disk, whilst $f_{\rm rem} = 0.3$ in the bulge and 0.15 in the
disk. The values of both $f_{\rm gas}$ and $f_{\rm rem}$ directly affect the
computed $M/L_R$ for each population as there is no light contribution from the
remnants or the gas. They therefore affect the way in which changes to the
stellar mass function translate into changes in the overall density
normalisation of a given component. Also, the larger the value of $f_{\rm rem}$
then the greater the microlensing contribution from remnant lenses at the
expense of stellar lenses, whereas a larger value of $f_{\rm gas}$ reduces the
overall microlensing signal since it is at the expense of both the remnant
and stellar density contributions.

\subsection{The microlensing models}

The models we use to compute the pixel-lensing rate are presented in
Table~\ref{models}. Model~1 is the most conservative comprising a light disk
and light exponential bulge. Model~2 considers a heavy disk and heavy
exponential bulge. Model~3 examines the effect of a light exponential bulge
combined with a heavy disk. The purpose of Model~4 is to allow us
to explore the effect of the luminosity function on pixel-lensing
predictions. It assumes a disk luminosity function and heavy disk mass
function for both the bulge and disk populations, though the bulge has the same
gas and remnant fraction as for the other bulge models.
The resulting mass-to-light ratio is $M/L_R = 2.4$, slightly lower than the
heavy disk value. So, whilst the bulge mass function in Model~4 is similar to
the heavy bulge model, the bulge mass is closer to that of the light bulge
because of the low mass-to-light ratio of the disk-like stellar population. 
Finally, Model~5 considers a light disk and power-law bulge to allow a
comparison between different functional forms for the bulge.
   
\section{Microlensing yields} \label{predictions}

We base our rate calculations on the parameters of the Angstrom Project,
assuming a distance to M31 of 780~kpc \citep{stan98} and a foreground
extinction of $A_R = 0.17$ \citep{sch98}.
The Angstrom Project uses three telescopes, one with a 11~arcminute field of view, the
other two with 5~arcminute fields of view. The data is being
analysed using difference imaging techniques \citep{ala98}, in which an optimal
reference image is constructed from images taken under good seeing conditions
and then subtracted from each target image after they have been convolved to
match the seeing characteristics of the reference image. The resulting
difference images record only objects that have varied in their flux. For our
calculations we assume observations are made in the $R$ band and the reference
image has a seeing of 1~arcsec. The typical  $R$-band exposure time is set at
600~sec and the detector has a zero point of $R = 24.6$~mag. For a Gaussian
source point spread function (PSF), and assuming the noise within the PSF is
dominated by the galactic background, the optimal signal-to-noise ratio is
obtained by measuring the difference flux out to 71.5\% of the encircled energy
in the PSF, corresponding to a radius of 0.67~arcsec for our reference image.
For guidance, the dashed lines in Figure~\ref{sources} indicate the limiting
source magnitude
assuming the parameters above for the detection of a pixel-lensing event with
$A< 30$ and a background surface brightness $\mu_R = 20$~mag~arcsec$^{-2}$.

\begin{figure}
\includegraphics[scale=0.35,angle=270]{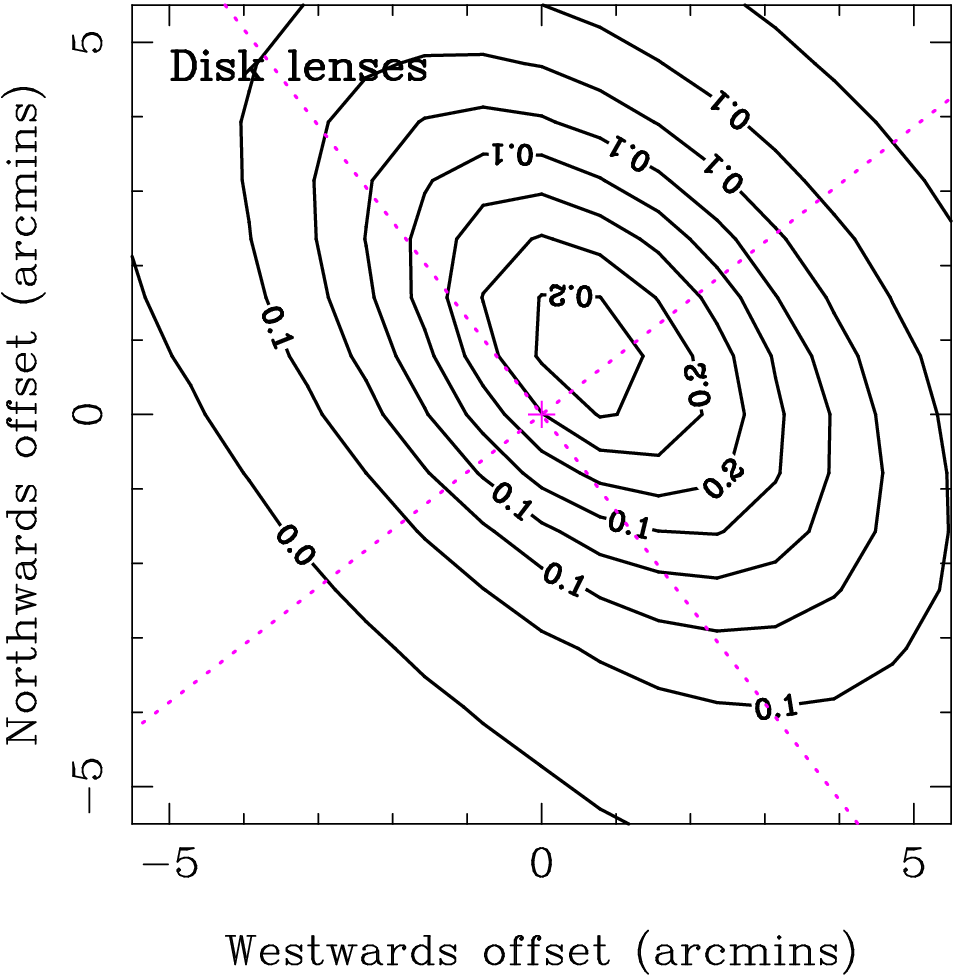}
\includegraphics[scale=0.35,angle=270]{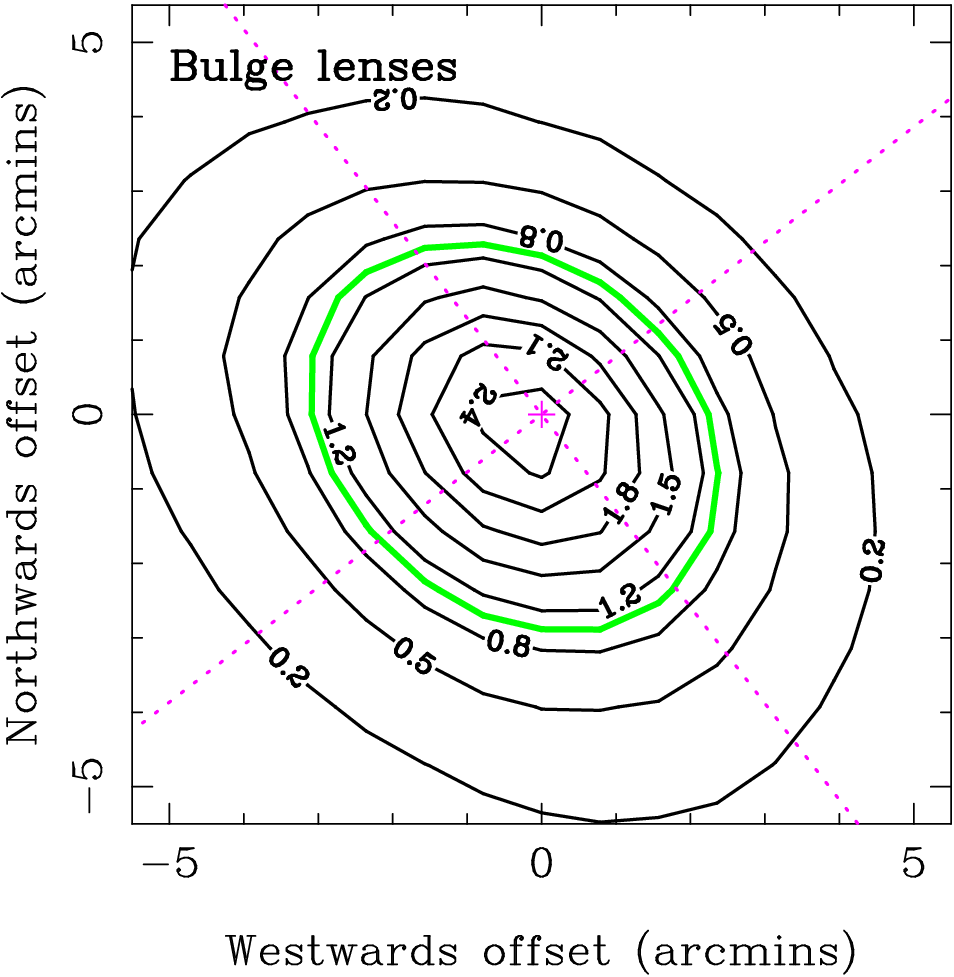}
\caption{The pixel lensing spatial distribution in model 1 for disk (top) and
bulge (bottom) lenses. Contours and lines are as for Figure~\ref{spatial}.}
\label{lenses}
\end{figure}

Figure~\ref{spatial} shows the spatial distribution of pixel-lensing events
with $1~\mbox{d}< \tv < 100~\mbox{d}$ for the five models in
Table~\ref{models}. The plots cover a field of view of 11~arcminutes centred on
the
bulge. The cross denotes the centre of M31 and the dotted lines show the
orientation on the sky of the major and minor axes of the M31 disk. The
contours show the expected number of events per square arcminute per year. For
all models the event distribution is strongly dominated by bulge lenses, as can
be seen in Figure~\ref{lenses}, which shows the spatial distribution separately
for disk and bulge lenses. The
number of events involving disk lenses is typically around 3 events per season,
 so it is around order of magnitude lower than the bulge rate even
for the heavy disk models. The spatial distributions in Figure~\ref{spatial}
therefore trace the bulge
structure and so are noticeably rotated away from the disk axes for all models.
Additionally, models 1--3 clearly show an excess of events towards the far
(South-Eastern) side of the M31 disk. For these models a significant fraction of
all events involve bulge lenses and disk sources, so we expect their
distribution to be skewed towards the far disk. Models 4 and 5 are dominated by
events involving bulge lenses and sources, so their distributions appear more
symmetric and more closely trace the bar profile.

\begin{figure}
\includegraphics[scale=0.35,angle=270]{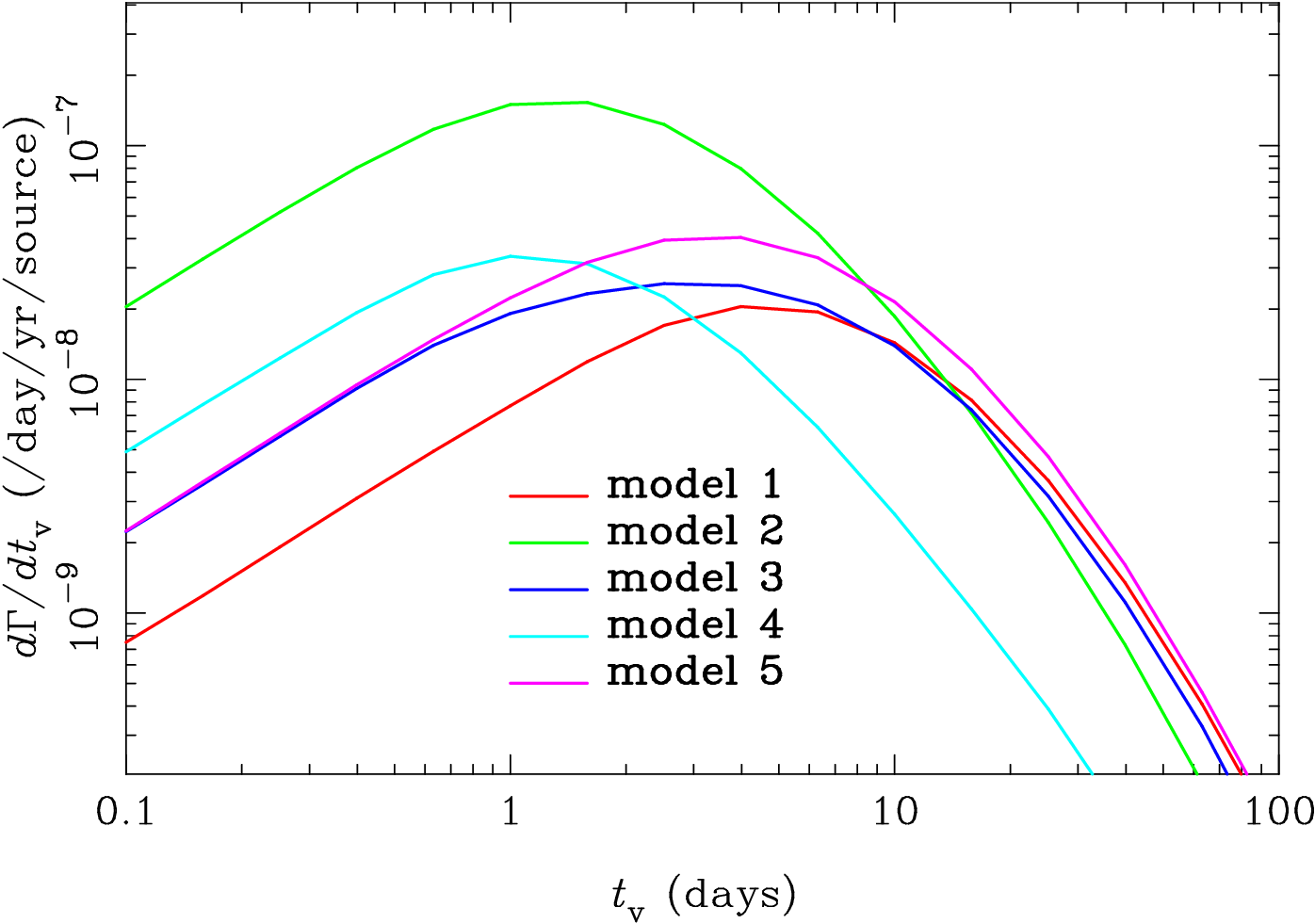}
\caption{The rate-weighted pixel-lensing timescale distribution for models 1--5
for an 11~arcminute field of view centred on the M31 bulge. The models are
most distinguishable for timescales $\tv < 10$~d.}
\label{times}
\end{figure}

\begin{table}
\caption{Event yields for the Angstrom Project. $\langle \tv \rangle$ refers
to the average visibility timescale for events in the range $1~\mbox{d} < \tv
< 100~\mbox{d}$. $N$ refers to the number of
events per five-month season over an 11~arcminute field centred on the M31 bulge
(as in Figure~\ref{spatial}). Angstrom is also using telescopes with 5~arcminute
fields, for which the expected yields are about half that quoted below. Yields
are also presented separately for the Eastern (Q1), Southern (Q2), Western (Q3)
and Northern (Q4) quadrants defined  in Figure~\ref{geom}.}
\label{yields}
\begin{tabular}{@{}cccccccc}
\hline
Model & $\langle \tv \rangle$(d) & \multicolumn{5}{c}{$N$(1-100~d)} &
$N$(1-10~d) \\
 & & Q1 & Q2 & Q3 & Q4 & Total & \\
\hline
1 & 16 & 8 & 8 & 6 & 5 & 27 & 13  \\
2 & 7 & 19 & 17 & 16 & 12 & 64 & 51 \\
3 & 14 & 8 & 8 & 7 & 5 & 28 & 15 \\
4 & 7 & 3 & 3 & 3 & 3 & 12 & 10 \\
5 & 13 & 14 & 13 & 13 & 10 &  50 & 28 \\
\hline
\end{tabular}
\end{table}

Figure~\ref{times} shows the timescale distributions for models 1--5. The
separation between the model predictions is most evident at timescales $\tv <
10$~days, emphasising the need for short-timescale sensitivity. In general the
heavier models produce a higher rate of events and peak at shorter durations.
The rate for all models tails off sharply for $\tv > 100$~days.

Table~\ref{yields} shows the average duration
and the number of events with visibility timescales between 1 and 100~d
expected from a five-month observing season for an 11~arcminute field centred on
the M31 bulge. The average duration lies between 7 and 16 days, with heavier
bulge models giving rise to shorter events. The shorter durations arise because
the
mass function for the heavier models extends into the low-mass brown dwarf
regime, and also because the velocity dispersion for the heavier models is
larger. 

Excluding model~4, the total
number of events with durations between 1 and 100~d ranges from 27 up to 64,
with the larger yields coming from the heavier, brown-dwarf rich  models. The
number of short-duration
events ($1< \tv< 10$~d) is also shown in Table~\ref{yields}, ranging from 13
events per season for the lighter models through to 50 events per season for the
heavy exponential bulge. 
For the lighter models about half of all events have
durations
below 10 days, whilst for the heavier models they comprise up to $80\%$ of all
events.

Model~4 assumes a disk luminosity function and heavy disk mass
function for the bulge stars. Observations clearly exclude such a luminosity
function for the M31 bulge, but the purpose of
this hybrid model is simply to assess the affect of changes to the luminosity
function on pixel-lensing predictions. It is evident from Figure~\ref{spatial}
and
Table~\ref{yields} that the yield from this model is significantly less than
for the others; it is a factor two below the conservative model~1.
Naively this seems a very surprising result given that the bulge sources are
brighter for model~4. Equation~(\ref{detect}) indicates that brighter
sources can be detected at larger impact parameters and therefore the rate
per source is larger from equation~(\ref{dr}). However, the low mass-to-light
ratio of the disk-like stars means that we must adopt a brown-dwarf dominated
mass function to give a bulge mass which is compatible with at least the lower
end of the preferred M31 bulge mass estimates, whilst satisfying the observed
M31 surface brightness. Even with a brown-dwarf dominated mass function, a
heavier bulge would be too bright to be consistent with surface brightness
observations. The combination of low-mass lenses and bright giant stars
typically gives rise to finite source effects which dilute the overall
microlensing magnification. For a $0.01~\sm$ lens lying at a typical distance of
1~kpc in front of the source, a lens--source angular separation of less than
$0.04~\mu$arcsec is required to produce a magnification $A>10$. At the distance
of M31 this corresponds to a separation of just 6~R$_{\sun}$, so finite source
effects are dominant at these lens masses, resulting in a significant
suppression of the microlensing rate for this model. This model is obviously
rather extreme, but the basic point is that the boost in the microlensing rate
per source due to brighter stars is cancelled out by the lower density
normalisation required to satisfy surface brightness constraints. So, even if
there are no constraints on the permitted mass of the bulge, a boost in the
number of bright stars does not result in a gain in the overall microlensing
rate.

The number of events expected within each quadrant Q1--4
defined in
Figure~\ref{geom} is also given in Table~\ref{yields}. For models 1--3 the ratio
of the number of events in each quadrant is about 1.6:1.5:1.2:1 (Q1:Q2:Q3:Q4),
whilst for model~4 it is 1.2:1:1.2:1 and for model~5 (the power-law bulge)
it is 1.5:1.3:1.3:1. The number of
events in quadrants Q1 and Q2 generally outnumber those in Q3 and Q4 because
the M31 disk inclination skews the distribution of bulge lens--disk source
events towards the far disk. This along with the offset of the barred bulge
away from the disk axis introduces an enhanced asymmetry between events at
either end of the bar in Q1 and Q3, giving rise overall to a four-way
asymmetry. The asymmetry is much less evident for model~4 because the
source luminosity and lens mass functions are the same for the disk and bulge,
so the rate of bulge lens--disk source events is similar to the rate of disk
lens--bulge source events.
For the other models, the strength of this four-way asymmetry depends upon the
bar offset, the degree
of bar prolongation, and the mass of the bar, so it is a good signature with
which to probe the bar geometry. For the models 1--3, a
sample of around 28 events would be needed in Q4 in order to detect a
difference with Q1 at the $95\%$ confidence level. Such a difference should
therefore be detectable within three to six seasons, depending on the model.
However, a much more
efficient strategy is simply to undertake a
likelihood analysis of the positions and timescales of the entire sample in
order to constrain the bar parameters.

One factor which can affect the observed event rate is extinction. We have allowed for both foreground and internal extinction in calibrating our M31 models (see Section~\ref{lf}). We have also allowed for foreground extinction
in computing the M31 distance modulus, but we have ignored internal M31 extinction for our rate calculations. \citet{hol05} have obtained internal extinction estimates for several nearby spiral galaxies by studying background galaxy counts. They find that extinction estimates are relatively independent of the galaxy inclination angle, which suggests that the dust is confined to a thin sheet in the disk plane. In this case it is easy to quantify the affect of extinction on microlensing rates. The half of the source population which lies in front of the dust screen is unaffected by the dust. The other half is dimmed by a factor $10^{-0.4 \, A_R} = 0.74$ for an internal extinction of $A_R = 0.33$~mag. From equation~(\ref{detect}) we see that, for a fixed galaxy brightness, a $26\%$ reduction in source flux demands a similar reduction in the maximum impact parameter $\ut$ in order for the microlensed source to be detected. Since from equation~(\ref{dr}) the microlensing rate scales with $\ut$ then an internal M31 extinction of $A_R = 0.33$ results in an overall reduction of $13\%$ in the observed rate.

\section{Discussion} \label{discuss}

The Angstrom
Project is using three telescopes to conduct a high time resolution survey of
the bulge of the Andromeda Galaxy (M31). The principal aim of the survey is to
detect microlensing events with durations above one day due to low mass stars
and brown dwarfs in the bulge of M31. 

We have shown in this study that M31 presents a ripe
target for a high time resolution microlensing survey. Using simple analytic
models for the disk and bulge of M31 we have shown that a survey like Angstrom
should detect around 27 events per season over an 11~arcminute field of
view for a standard stellar mass function which is truncated at the hydrogen
burning limit. Half of these events have durations below 10 days. If the stellar
mass function contains a substantial brown-dwarf population the expected yield
raises to around 64 events per season, with some $80\%$ having durations less
than 10
days.

The underlying mass function of the lenses can be probed through the average
duration of detected events, as well as through the number of events.
Mass functions with dominant brown dwarf populations produce shorter events but
also necessarily imply a larger mass-to-light ratio and so heavier bulge or disk
mass, giving rise to more events.

We find that the events are dominated by bulge lenses and that their spatial
distribution provides a good tracer of the bulge geometry. In particular, if
the bulge is barred and orientated away from the disk major axis, as
suggested by surface brightness measurements, then this should be evident in
the spatial distribution of the events. The importance of this is that
microlensing directly traces the underlying mass distribution, rather than the
light distribution, and so provides a completely independent probe of the
underlying bar geometry. Lastly, the combination of an offset barred bulge,
a highly inclined disk, and different stellar luminosity functions in the disk
 and bulge is predicted to give rise to a four-way asymmetry in the
number of events. Different numbers of events are expected in each of the four
quadrants defined by the major and minor axes of the M31 disk.

Aside from probing the M31 bulge structure and mass function, the high time
resolution data from the Angstrom Project should allow for a more efficient
detection of exotic microlensing phenomena, such as
binary lensing events \citep{bal01}, than is possible with current surveys. It
will also be uniquely sensitive to variable stars and transients which undergo
outbursts on
timescales of less than 10 days.

\section*{Acknowledgements}

We thank Maurizio Salaris for running the synthetic colour magnitude simulations which were employed in this work.
The work of EK is supported by an Advanced
Fellowship from the Particle
Physics and Astronomy Research Council. MJD and JD are supported
by PhD studentships from the Particle
Physics and Astronomy Research Council. Work by AG is supported by NSF grant
02-01266. Work by CH is supported by the Astrophysical Research Center
for the Structure and Evolution of the Cosmos (ARCSEC) of the
Korea Science and Engineering Foundation, through the Science
Research Center Program. BGP is supported by the Ministry of
Science and Technology of Korea.

\end{document}